# Measurement of the impact of turbulence anisoplanatism on precision free-space optical time transfer

W. C. Swann,[1,*] M. I. Bodine,[1] I. Khader,[1] J.-D. Deschênes,[2,3] E. Baumann,[1] L. C. Sinclair,[1] and N. R. Newbury[1]

[1] National Institute of Standards and Technology, 325 Broadway, Boulder, CO, 80305, USA
[2] Université Laval, 2325 Rue de l'Université, Québec, QC, G1V 0A6, Canada
[3] Octosig Consulting, Québec, QC, G1V 0A6, Canada
*Corresponding author: william.swann@nist.gov

Future free-space optical clock networks will require optical links for time and frequency transfer. In many potential realizations of these networks, these links will extend over long distances and will span moving platforms, e.g. ground-to-air or ground-to-satellite. In these cases, the transverse platform motion coupled with spatial variations in atmospheric optical turbulence will lead to a breakdown in the time-of-flight reciprocity upon which optical two-way time-frequency transfer is based. Here, we report experimental measurements of this effect by use of comb-based optical two-way time-frequency transfer over two spatially separated optical links. We find only a modest degradation in the time synchronization and frequency syntonization between two sites, in good agreement with theory. Based on this agreement, we can extrapolate this 2-km result to longer distances, finding only a few-femtosecond timing noise increase due to turbulence for a link from ground to a mid-earth orbit satellite.

# 1. Introduction

With the continued improvements in optical clocks and oscillators, there has been recent parallel development of free-space optical links to connect these clocks [1–10] toward the goal of future extended free-space networks for applications ranging from fundamental science to precision navigation and timing. To achieve the highest performance, these links must rely on optical signals rather than RF signals because of the higher available bandwidth. They must also rely heavily on the reciprocity, or equal time-of-flight, inherent in optical bi-directional single-mode links even across highly turbulent air [11] in order to cancel out variations in the time-of-flight. Recent comb-based optical two-way time-frequency transfer (O-TWTFT) has verified this reciprocity down to levels of 100 attoseconds in time and levels of $10^{-19}$ or below in fractional frequency [7,9,12]. However, if the clock platforms are moving, this reciprocity will break down. This is true both for motion along the line-of-slight and transverse to the link connecting the clocks. For line-of-sight motion, a newly developed implementation of comb-based O-TWTFT cancels the effects of relative velocity to below 100's of attoseconds [13,14]. In the present work, we explore the effects of motion traverse to the link experimentally and compare the results to previous theoretical discussions [15,16].

Here, we use a comb-based O-TWTFT system, which is capable of measuring time-of-flights across a turbulent atmosphere with sub-femtosecond timing precision. To mimic the beam path displacement caused by transverse motion, we implement the comb-based O-TWTFT across two displaced counterpropagating optical links rather than a single bidirectional link. Our results compare well with theory [15,16] out to averaging times of 10 to 100 s without invoking any free parameters and indicate that the degradation in optical time-frequency transfer due to turbulence-induced reciprocity breakdown should be negligible. At longer averaging times we do see a disagreement with theory, likely resulting from displacement of the terminals at our optical link ends, as explained in Section 4.

The cause of reciprocity breakdown with transverse motion is illustrated in Fig. 1. In establishing time transfer to an orbiting satellite or other rapidly moving platform, rapid transverse platform motion combined with the finite speed of light results in the uplink and downlink paths passing through the turbulent atmosphere being physically separated by a point-ahead angle, $\phi_{\text{PAA}}$. As this angle increases the two beam paths increasingly sample different turbulence volumes, resulting in increasingly larger phase variations between the paths. At separations beyond the isoplanatic angle $\phi_{\text{iso}}$, defined as the solid angle region within which the phasefront distortions are below 1 radian, the two beam paths become increasingly decorrelated and the turbulence-induced non-reciprocity, or anisoplanitism becomes apparent. The resulting non-reciprocal time-of-flight for the uplink and downlink paths is determined by this decorrelation between the two paths. Either real wind or a "pseudowind" given by a satellite slewing across the sky effectively drives the turbulence across the separated paths, converting spatial turbulence into a time-varying non-reciprocal time-of-flight. Critically, this non-reciprocal portion of the total time-of-flight would be indistinguishable from an actual clock time-frequency offset in a two-way comparison and thus limit both accuracy and stability across a network.

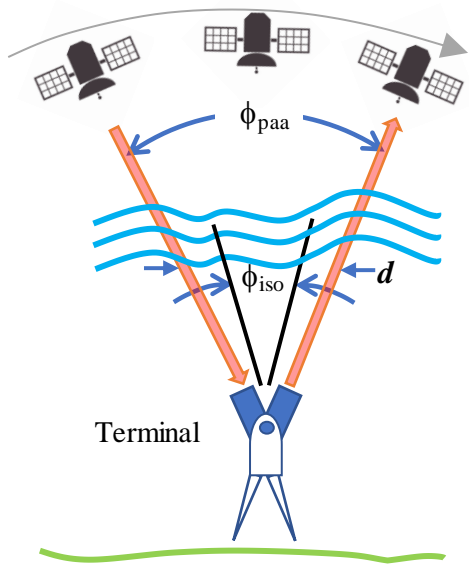

Fig. 1. Optical two-way time-frequency transfer to a satellite. The red arrows indicate the downlink and uplink paths. $\phi_{PAA}$ is the point-ahead angle and $\phi_{iso}$ denotes the isoplanatic angle, defined as the angular region within which phasefront distortions are below one radian. The separation between the downlink and uplink paths is given by $d$, which varies with altitude.

The effects of turbulence anisoplanitism have been well studied in the fields of both astronomical imaging [17,18] and free-space optical communications [19–23]. Anisoplanatism appears in O-TWTFT much as it does in astronomical imaging: the same turbulence that produces spatial distortions across an astronomical image will also produce differential time-of-flight variations across the uplink and downlink paths. Additionally, the impact of turbulence anisoplanitism on two-way time-frequency transfer has been analyzed in two recent theoretical efforts. Robert et al. [15] numerically model the effects of turbulence on both the phase noise and heterodyne efficiency of an optical link to a satellite in low-earth orbit, finding a fractional frequency stability below $1\times10^{-17}$ at 1-second averaging time should be achievable despite turbulence. Belmonte et al. [16] present a filter-based integral model for the non-reciprocity in the time-of-flight to a satellite in mid-earth orbit, showing a time-transfer standard deviation of only a few fs over a averaging times between 0.1 s and 1000 s. Here, through experiment, we quantify the effect of anisoplanatism over a km-scale horizontal link, compare the measured results to the theory of Ref. [16], and extrapolate to satellite-based links.

## 2. Theoretical Background

In two-way transfer between two clocks, one at site A and one at site B, two counter-propagating optical timing signals are transmitted, from A to B and from B to A. A comparison of their arrival times according to the local clocks at A and B yields the relative clock timing between the sites, independent of the varying time-of-flight, provided it is reciprocal or equal over the two paths. Specifically, if $T_{A \to B}$ is the time-of-flight from Site A to B and $T_{B \to A}$ is the time-of-flight from Site B to A, then the time-of-flight drops out of the two-way comparison provided $T_{A \to B} = T_{B \to A}$. Alternatively, if the difference in time-of-flights is known and calculable, it can also be removed from the two-way comparison, as in Refs. [13,14]. However, any uncontrolled non-reciprocity in the time-of-flight, $T_{NR} = T_{A \to B} - T_{B \to A}$ will appear as timing noise. Here, we probe that non-reciprocity by operating comb-based O-TWTFT with a common clock serving both sites. We purposefully break the bidirectionality of the optical link by connecting A to B across one path and B to A across a second path, displaced by a distance $d(z)$, where $z$ is the along-path coordinate. We then measure the time-dependence of $T_{NR}$ due to turbulence at different path separations $d$.

The question then becomes, what is the structure of this non-reciprocal time-of-flight, $T_{NR}$, which arises from the turbulence-induced anisoplanatism. Atmospheric optical turbulence can be described by turbulent eddies of varying size and refractive index. The spatial structure of these refractive index variations (which have their root in small, local temperature variations throughout the atmosphere) is described by the modified von Karman turbulence spectrum [24],

$$\Phi_n(\kappa) = 0.033 C_n^2 \kappa^{-11/3} F(\kappa) \tag{1}$$

where $C_n^2$ is the refractive index structure parameter, $\kappa$ is the scalar spatial frequency and $F(\kappa) = \left(1 + (1/\kappa L_0)^2\right)^{-11/6} \exp(-0.029 \kappa^2 l_0^2)$ describes the roll-off at the inner scale, $l_0$, and outer scale, $L_0$. Under typical atmospheric conditions $l_0$ is ~ 2 to 3 mm in size near the ground, increasing to several centimeters at altitudes above 10 km [25]. The outer scale parameter $L_0$, which represents the largest eddies, is not well defined but is often taken as on the order of 100 m.

In the "frozen turbulence" model the atmospheric turbulence itself is considered static. Wind (either real or pseudowind) of velocity transverse to the link having magnitude $V$ drives this frozen turbulence across the link, and through the relationship $\kappa = 2\pi f / V$ spatial turbulence variations are mapped into the frequency domain, where $f$ is Fourier frequency. Eq. (6) of Ref. [16] derives the power spectral density (PSD) for $T_{NR}$ for a link of length $L$,

$$S_T(f) = \frac{0.26 \pi^2}{4 c^2} f^{-8/3} \int_0^L dz \left(\frac{2\pi}{V(z)}\right)^{-5/3} C_n^2(z) F\left(\frac{2\pi f}{V(z)}\right) \left\{1 - J_0\left(\frac{2\pi f d(z)}{V(z)}\right)\right\} \tag{2}$$

where $z$ is along the link, $c$ is the speed-of-light, and $J_0$ is a Bessel functions of the first kind of order 0. Both the path separation $d$ and the refractive index structure parameter $C_n^2$ can vary over the path, as will be the case for a ground to satellite link (see Fig. 1). Note that we have included an additional factor of 4 in the denominator of Equation (2) compared to the derivation of Ref. [16]; this allows for a more direct comparison with our measurements as the non-reciprocal time-of-flight enters the clock offset computation as $T_{NR}/2$.

The term in curly brackets in Equation (2) serves as a "filter" to roll off the power spectrum slope for the $T_{NR}$ PSD from $f^{-8/3}$ to $f^{-2/3}$ for $f \leq 0.3 V/d$. Roughly stated, the turbulence-induced time-of-flight between the two paths becomes correlated at low frequencies as the turbulence eddies extend across the path separation. When $d=0$, i.e. a fully reciprocal path, $J_0 \to 1$ and $T_{NR} \to 0$ and $S_T(f) \to 0$, as expected.

If the term in curly brackets is excluded, and after multiplication by 2, Equation (2) is the PSD for the one-way time-of-flight $T_{link}$ which is responsible for significant time/frequency noise in any one-way time-frequency transfer. It has been discussed and compared to experiment previously in Refs. [1,26]. In between the cutoff frequencies set by the inner and outer scale, this PSD is predicted from Eq. (2) to follow a $f^{-8/3}$ rolloff. Experimentally, Ref. [26] found reasonable agreement with a $f^{-7/3}$ rolloff. However, Ref. [26] found this $f^{-7/3}$ scaling continued to very low Fourier frequencies, with no evidence of a low-frequency rolloff due to the outer scale. The lack of an outer scale is attributed either to a much larger outer scale than expected or, more likely, to actual temporal variations in the turbulence that are ignored in the frozen-turbulence model.

Equation (2) lends itself to straightforward modelling for comparison with field measurements. Here, in the experiment, the horizontal path is assumed to have uniform turbulence and wind speed. In this case, with knowledge of the ambient wind speed $V$, turbulence strength $C_n^2$, and the path separation, $d(z)$, we can compare the model and measurement directly with no free parameters. Figure 2 shows the calculated PSD as the path separation is changed, while all other parameters remain fixed. The roll-off of the spectral slope from $f^{-8/3}$ to $f^{-2/3}$ occurs at $f = 0.3V/d$.

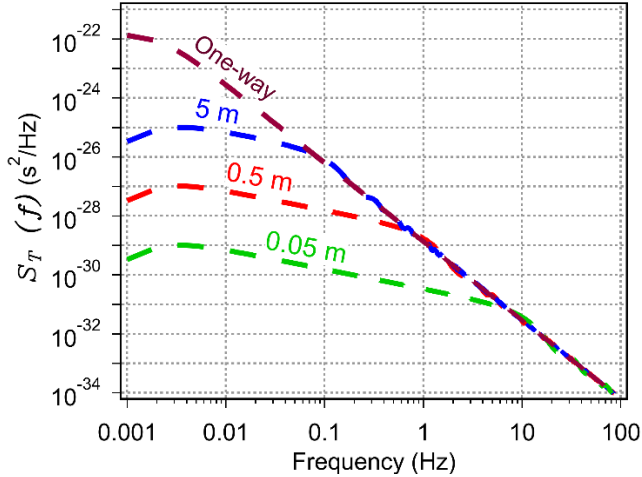

Fig. 2. Modelled $T_{NR}$ power spectral density (PSD) given by Eqn. 2 for path separations $d = 5$m (blue), 0.5m (red) and 0.05m (green). The slope decreases from $f^{-8/3}$ to $f^{-2/3}$ at lower Fourier frequencies because of correlated turbulence across the two paths. The roll-off below 0.005 Hz results from a turbulence outer scale $L_0$ of 100 m. (High-frequency roll-off from the inner scale $l_0$ and from aperture averaging fall outside the range of the plot.) Also shown on the plot is the one-way time-of-flight PSD (dark red).

## 3. Experimental setup

We use an existing comb-based O-TWTFT setup as illustrated in Fig. 3 and described in more detail in Refs. [7,10]. As illustrated in Fig. 3, at each end of the link, a site consists of a frequency comb phase-locked to a common cavity-stabilized laser, a pair of transmit and receive terminals [27], an optical transceiver to perform optical heterodyne timing measurements and the necessary digital signal processing infrastructure. Use of a common cavity-stabilized laser removes any true clock offsets to isolate the potential clock error due to a non-zero $T_{NR}$.

The frequency comb at each site provides the pulse train (or timing signal) that is sent across the link; it also serves as a local oscillator for measurement of the arrival time of the remote comb signal. The pulse train launched from each fully-stabilized frequency comb is centered at ~1560 nm and has a bandwidth of ~12 nm [28]. The combs operate at a nominal repetition frequency of 200 MHz and with an offset in repetition frequencies of ~2 kHz between them as required by the linear optical sampling technique [10,29] used to interrogate the received pulse train with the local frequency comb.

In normal two-way operation, only one free-space optical terminal resides at each site and operates bi-directionally, both transmitting the local comb and receiving the remote comb signal. Here, as shown in Fig. 3, optical circulators separate the transmitted and received signals for two terminals at each site. The separation $d$ between the terminals can be varied from 0.045 m to 1.1 m. A 0.5 m diameter mirror at 1 km distance from the terminals folds the path so that the two clock sites are adjacent for use of the common cavity-stabilized laser. Because the path is folded, the path separation varies along the path, becoming zero at the fold mirror. (The use of a folded path also introduces a second filter function in the non-reciprocal time-of-flight but with a cutoff frequency too low to be measured here). Both terminal pairs are aimed at the fold mirror through laboratory windows that are closed to guarantee no excess turbulence caused by air mixing through open windows.

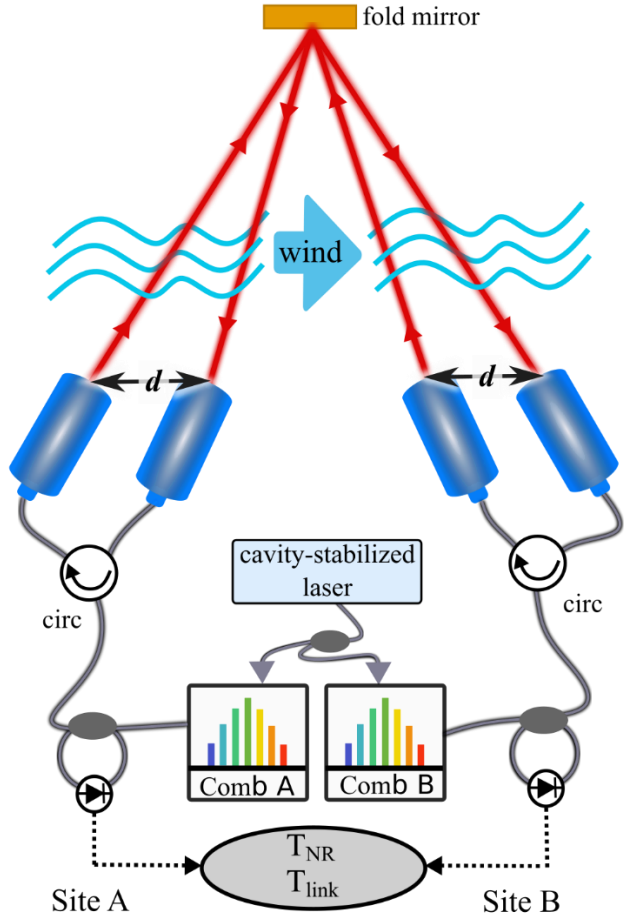

Fig. 3. Experimental setup. Time-of-flight reciprocity is broken by separating the timing signals onto two unidirectional paths. Transfer from Site A to Site B is via the outer path. Transfer from Site B to Site A is via the inner path. The separation $d$ between the two paths at the terminals is varied. The measurement returns both $T_{link}$ and $T_{NR}$. Det: balanced detector. Circ: circulator.

The non-reciprocal time-of-flight, $T_{NR}$, and the average time-of-flight, $T_{link}$ are extracted from the O-TWTFT at an update rate of 2 kHz by use of the equations given in Refs. [13,14]. The wind speed is measured using a 3-D anemometer near the location of the terminals. Path-averaged $C_n^2$ values are extracted using a gamma-gamma fit to the received signal's atmospheric turbulence-induced amplitude scintillation [30,31]. Using the received signal to estimate $C_n^2$ assures that the value returned truly reflects the turbulence conditions over the link.

It should be noted that although driven by experimental constraints, a 1-meter path separation is not an unreasonable choice for exploring anisoplanatism effects on time transfer to a satellite for the following reason. The point ahead angle $\phi_{PAA}$ to a satellite in mid-earth orbit (MEO) is 50 microradians. This gives a path separation at an altitude of 10 to 20 km (a typical altitude for upper-atmosphere turbulence) of ~0.5 to 1 meter.

## 4. Results

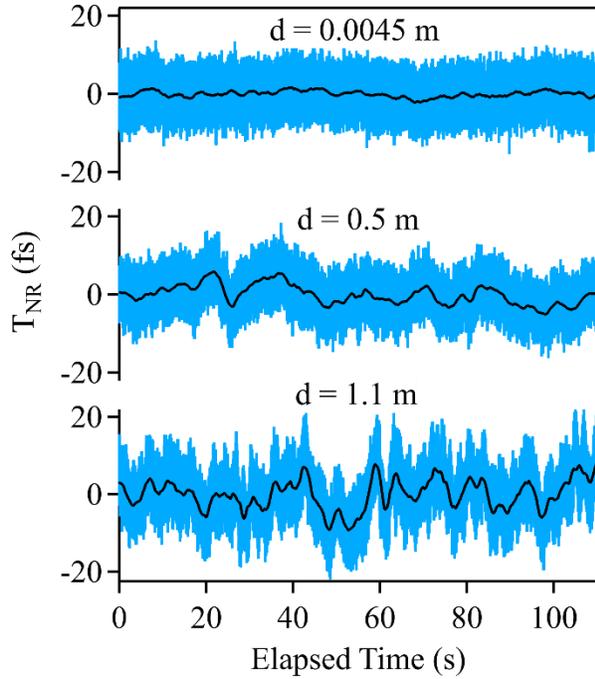

Fig. 4. Time series of $T_{NR}$ for separations of 0.045 m (top), 0.5 m (middle) and 1.1 m (bottom) at the at the full 2 kHz measurement rate (blue) and smoothed to 0.5 Hz (black). An increase in low frequency noise due to non-reciprocity is clearly seen for both $d$=0.5-m and $d$=1.1-m.

We measured the non-reciprocal time-of-flight $T_{NR}$ over our 2-km link for terminal separations, $d$, between 0.045 m and 1.1 m. The measurements took place over the course of several weeks. Winds were low, varying between 0.4 and 4 m/s over this time. $C_n^2$ varied between ~ $3\times10^{-15}$ m$^{-2/3}$ and $1.5\times10^{-14}$ m$^{-2/3}$. Figure 4 shows a time series of $T_{NR}$ for separations of 0.045 m, 0.5 m, and 1.1 m. In all cases, there is a ~3 fs standard deviation at the full 2 kHz sampling rate, caused by the system noise floor. However, at the 0.5 and 1.1-m separation, there is additional low frequency noise having standard deviations of ~3 fs and ~4 fs, respectively, at a 0.5 Hz sampling rate, caused by the turbulence variation between the two paths.

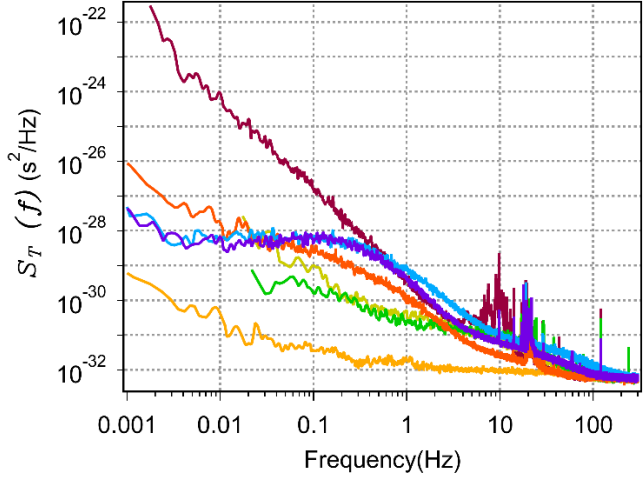

Fig. 5. PSD of the measured $T_{NR}$ for path separations, $d$, of 1.1m (violet), 0.75m (blue), 0.5m (red), 0.06m (yellow), and 0.045m (green), and a fully reciprocal link (orange). Shown in dark red is the PSD for the one-way $T_{link}$. The spikes between 10 Hz and 30 Hz are due to differential acoustic and seismic noise on the transmit/receive terminals and do not appear on the fully reciprocal link.

Figure 5 shows the measured $T_{NR}$ PSDs for five terminal separations $d$ between 0.045m and 1.1m. We observe good qualitative agreement with Fig. 2. In particular, over the spectral band between 0.1 Hz and 4 Hz, the PSD increases with separation as expected. The 0.75 m data (blue trace) were taken at a higher wind speed and higher $C_n^2$ than the other data, which shifts the PSD up as predicted. At frequencies above 60 Hz, turbulence effects fall to the measurement noise floor (orange trace). The PSD for the averaged time-of-flight $T_{link}$ for the 0.5 m separation is also shown and follows an $f^{-2.6}$ slope between 0.005 and 4 Hz. This is between the $f^{-7/3}$ and $f^{-8/3}$ slopes expected from Ref. [26] and theory, respectively.

One feature seen in Figure 5 is not in agreement with theory: the PSDs show no signs of an outer turbulence scale $L_0$ roll-off but rather continue to increase at low frequencies. We attribute this increase to thermally-induced mechanical displacement in the terminal positions, specifically micron-level shift of the inner terminal pair relative to the outer pair (see Fig. 3) as would be caused by room temperature changes or solar loading causing distortion of the structure on which the terminals rest. This type of movement is consistent with the Aluminum substructure and ambient temperature changes. Moreover, the terminal gimbals did require sub-milliradian adjustment throughout the measurement to maintain pointing, which substantiates the small relative physical movement of the terminals. This effect is especially evident in the 0.045 m and 0.06 m traces, which had particularly non-robust mechanical arrangements. Note that for the case of separate transmit and receive terminals, the above indicates that at very low Fourier frequencies the exact physical configuration of the terminals will dominate over any turbulence-induced effects. While Ref. [26] also reports an absence of an outer scale roll-off potentially due to temporal variations not captured by the frozen turbulence model, such a temporal effect would be common to the separated paths.

Figure 6 quantitatively compares the PSD at the $d=0.5$ m separation to Eqn. (2) using the measured values of $V = 0.6 \pm 0.4$ m/s and $C_n^2 = 5.6 \pm 1.6 \times 10^{-15}$ m$^{-2/3}$. The model and experiment agree quantitatively below ~4 Hz, except for the increase at very low Fourier frequencies of a few mHz discussed above. (Note that the outer-scale roll-off is not included in this model comparison.) Similar agreement between the model and measurements are seen for the other separations.

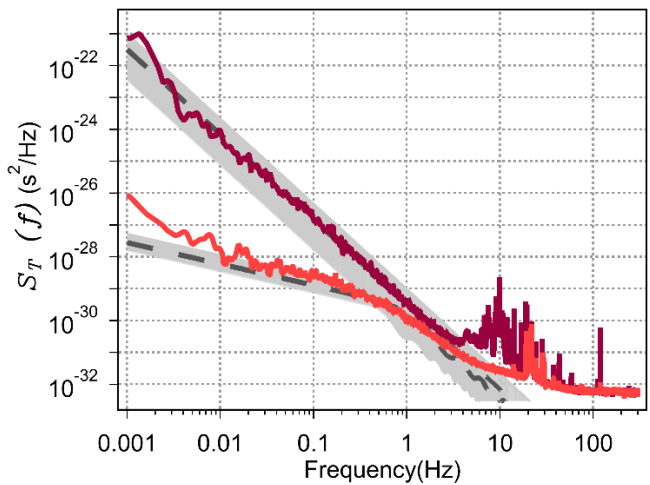

Fig. 6. Measured PSD for $T_{NR}$ (red) and $T_{link}$ (dark red) for a path separation of $d = 0.5$ m, as compared with the model (grey dashed line), including the range due to varying wind speed and turbulence (shaded grey region). Agreement between the measurement and model is reasonable.

Figure 7 shows the calculated time deviation from the data at the five separations for both the non-reciprocal two-way link and for the one-way link. Since a common clock was used, this time deviation for the two-way link is the excess noise from the non-reciprocity. In general, the smaller separations show a lower time deviation for the two-way data between 0.1 and 10 s, mirroring the trend seen in the PSDs. The time deviations remain below 10 fs between 1 ms and 1000 s. All two-way time deviations show an increase from their minima at longer averaging times, reflecting the continued increase in the PSD at low frequencies, as discussed above.

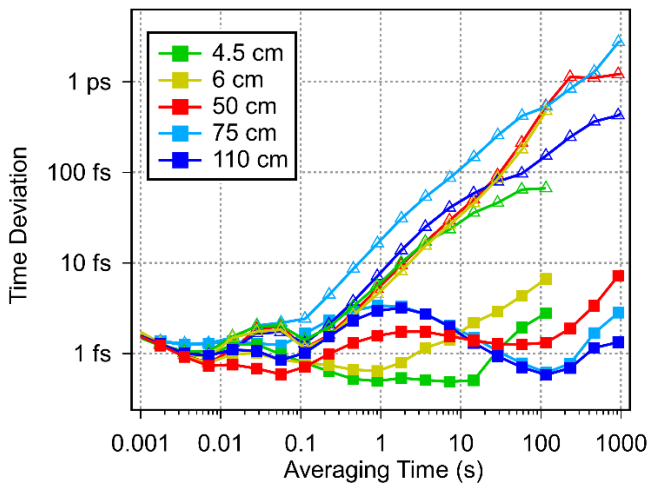

Fig. 7. Time deviation vs averaging time. The time deviation corresponding to the one-way transfer, i.e. of $T_{link}$, (open triangles) increases roughly linearly with time after about 0.1 s, whereas the time deviation for two-way transfer, i.e. of $T_{NR}$, (solid squares) remains roughly flat at below 10 fs.

## 5. Discussion

The results presented here show agreement between the measurements and model at all but very low Fourier frequencies (or long averaging times), giving confidence in the model's ability to predict $T_{NR}$ under differing path separation and atmospheric conditions. We can therefore model the turbulence-induced time deviation for transfer to a satellite in MEO. For the model, the satellite elevation is 45 degrees above the horizon, the ground-level $C_n^2$ is $6\times10^{-15}$ m$^{-2/3}$, and the ground-level wind velocity is 0.6 m/s. The satellite has an altitude of 9000 km, a velocity $V$ of 5.4 km/s, and a slew rate of 400 µrad/s giving a point-ahead angle $\phi_{PAA}$ of 36 µrad. Note that these conditions are similar to those presented in [16]. The model itself includes variation of $C_n^2$ vs altitude given by the widely-accepted Hufnagle-Valley turbulence model and upper-altitude winds as given by the Bufton wind model [24]. For the model to be well-behaved and give realistic results at short averaging times, we include in the model our typical measured white noise floor of $5.7\times10^{-33}$ s$^2$/Hz.

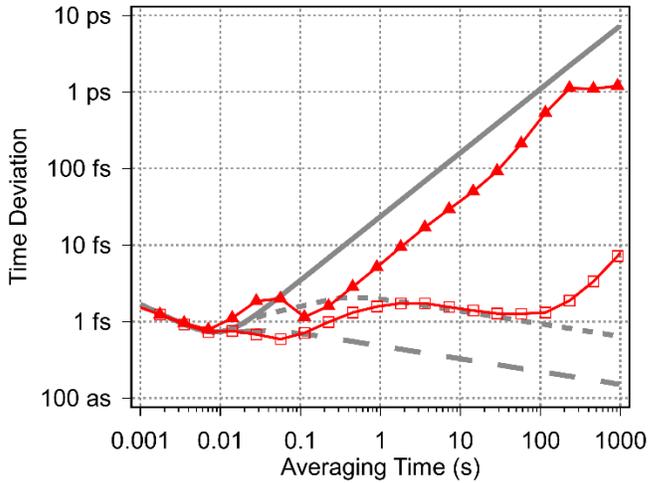

Fig. 8. Modelled time deviation vs averaging time for transfer to a satellite at a ground terminal separation of 0.5 m (gray short dashes) and 0 m (gray long dashes). The time deviation for a one-way satellite link is also shown (gray solid). The measured time deviation over the 2-km link at a 0.5 m- terminal separation is also shown for both the two-way (red open squares) and one-way (red triangles) configurations.

Figure 8 shows the model results for ground-level terminal separations of both 0 m and 0.5 m as well as for a one-way link. Even at the 0.5-m separation (gray, short dashes), the time deviation remains below ~2 fs over all averaging times, similar to the ~5 fs at 1000 s presented in Fig. 2 of [16] as expected given the similar conditions and including our measurement noise floor. If the same terminal is used for both transmit and receive (terminal separation of 0 m) the time deviation is reduced by about 6 dB, demonstrating that the ground-level atmospheric conditions are responsible for a substantial portion of the timing noise if separate, displaced ground terminals are used. As shown in the Figure 8, the time deviation for a one-way link reaches ~6 ps at 1000 s, ~ 33 dB higher than for the non-reciprocal two-way link.

The measured time deviation at our 0.5 m terminal separation over the 2 km horizontal link is included in Fig. 8 to show the remarkable agreement with the modelled path to MEO, out to an averaging time of 100 s. This general agreement is expected for the following reason: Our entire 2 km measured path is through low-altitude turbulence, whereas the modeled low altitude turbulence falls off rapidly with altitude, with $C_n^2$ dropping from $\sim 1\times 10^{-14}$ m$^{-2/3}$ at ground level to $\sim 1\times 10^{-16}$ m$^{-2/3}$ at 1 km altitude. The long path through upper altitude turbulence in the model approximately equals the turbulence over most of our relatively short 2 km near-ground path. Beyond 100-second averaging time, the data lies above the model due to differential displacement of the terminal pairs, as discussed above. Note that this increase arises from the difference between ideal model and physical reality, and thus could occur in any system that has separate, displaced transmit and receive terminals. Its suppression requires appropriate thermal and vibration control at the transmit/receive telescope.

## 6. Conclusion

We have presented measurement results for anisoplanatic non-reciprocal time-of-flight on time transfer across a 2 km link for varying path separations. We find good general agreement between the experimental data and model when comparing our results to a general model that predicts the non-reciprocal time-of-flight for varying atmospheric conditions and path separations. Extrapolation of these results to a ground-to-satellite link indicate that the effects of anisoplanitism and the point-ahead angle should not degrade the time deviation beyond a few femtoseconds for a

MEO orbit. We do note that the model prediction is based on the Hufnagle-Valley turbulence model and further quantitative study of upper-atmosphere turbulence would improve the model results.

**Acknowledgments**

This work was funded by the Defense Advanced Research Projects Agency (DARPA) PULSE program and the National Institute for Standards and Technology (NIST). We thank Mick Cermak for general technical assistance and Emily Hannah and Steve Jefferts for comments.